\documentstyle[preprint,aps,prl,psfig]{revtex}
\tighten
\draft
\begin{document}

\draft

\preprint{September 16, 1998}
\title{The specific heat of amorphous silica within the harmonic 
approximation}
\author{J\"urgen Horbach, Walter Kob, and Kurt Binder}
\address{Institut f\"ur Physik, Johannes Gutenberg-Universit\"at,
Staudinger Weg 7, D-55099 Mainz, Germany}
\maketitle

\begin{abstract}
We investigate to what extent the specific heat of amorphous silica can
be calculated within the harmonic approximation. For this we use
molecular dynamics computer simulations to calculate, for a simple
silica model (the BKS potential), the velocity autocorrelation function
and hence an effective density of states $g(\nu)$. We find that the
harmonic approximation is valid for temperatures below 300K but starts
to break down at higher temperatures. We show that in order to get a
reliable description of the low frequency part of $g(\nu)$, i.e. where
the boson peak is observed, it is essential to use large systems for
the simulations and small cooling rates with which the samples are
quenched. We find that the calculated specific heat is, at low
temperatures (below 50K), about a factor of two smaller than the
experimental one. In the temperature range 200K $\leq T \leq T_g$,
where $T_g=1450$K is the glass transition temperature, we find a very
good agreement between the theoretical specific heat and the
experimental one.

\end{abstract}


\section{Introduction}
\label{sec0}
\noindent
It is well known that many low-temperature properties of crystals, such
as the specific heat or the vibrational dynamics of the atoms, can be
calculated if the wave-vector dependence of the density of states (DOS)
is known. At higher temperatures the system will in general become
anharmonic and thus the validity of the harmonic approximation breaks
down. For {\it amorphous} systems at low temperatures the situation is
similar to the one of crystals. However, if the temperature is
increased not only anharmonic effects have to be taken into account but
also the relaxation dynamics of the system since the latter will in general
even take place at temperatures significantly below the glass
transition temperature $T_g$ and thus make the harmonic approximation
invalid. That the low temperature specific heat $C_V$ of silica can
indeed be calculated reliably from the DOS was demonstrated by Buchenau
{\it et al.} who used the DOS, as determined from neutron scattering,
to calculate $C_V$ between 5 and 20 Kelvin~\cite{buchenau86}. Very
recently Taraskin and Elliott presented the results of a computer
simulation in which they had calculated the specific heat of silica in
a similar temperature range~\cite{taraskin97b}. They found that the
calculated specific heat is smaller than the experimental one and
conjectured that the discrepancy might be due to the relatively small
size of their system.

What so far has not been investigated is to what extent the harmonic
approximation can be used to calculate the specific heat also at
higher temperatures, i.e. in the range 100~K $\leq T \leq T_g$, where
$T_g=1450$~K is the glass transition temperature. In the present work
we therefore present the results of a calculation of the specific
heat in this temperature range. For this we use the DOS obtained from
a large scale molecular dynamics computer simulation and compare the
so obtained results with the experimental values.

The rest of the paper is organized as follows: In the next section we
give the details of the model used and of the molecular dynamics
simulation. In Sec.~\ref{sec3} we present the results and summarize
these in the last section.

\section{Model and Details of the Simulation}
\label{sec1}

In this section we give the details of the simulations and describe
how the glass samples have been generated.

The silica model we are using in this simulation is the one proposed by
van Beest {\it et al.}~\cite{beest90}. In this model the interactions 
$\phi(r_{ij})$ between two ions $i,j$ that are separated by a
distance $r_{ij}$ is given by 

\begin{equation}
\phi(r_{ij})=\frac{q_i q_j e^2}{r_{ij}}+A_{ij}e^{-B_{ij}r_{ij}}-
\frac{C_{ij}}{r_{ij}^6}\quad .
\label{eq1}
\end{equation}

The values of the partial charges $q_i$, measured in units of $e$, as
well as the values of the constants $A_{ij}$, $B_{ij}$, and
$C_{ij}$ can be found in Refs.~\cite{beest90,vollmayr96b}. One of the
remarkable features of this potential is that it contains only 
{\it two}-body interactions, thus making it very
appealing from a computational point of view, since the evaluation of
computationally demanding three-body forces is avoided.  Despite its
relative simplicity, previous investigations have shown that this model
gives a realistic description of the static properties of silica
glass~\cite{vollmayr96b,vollmayr96a} (temperature dependence of the
density, structure factor, bond-angle distribution functions), and also
of the dynamical properties of silica
melts~\cite{horbach98a,horbach98b} (diffusion constants, viscosity,
intermediate scattering functions).

For the present simulation we use 8016 ions, in a rigid cubic box of
size 48.37\AA. This corresponds to a density of 2.37 g/cm$^3$ which is
close to the experimental value for amorphous silica and a pressure
which is a bit higher than the normal pressure. The reason for
choosing a system size that, in the field of simulations of supercooled
liquids and glasses, is relatively large, is that the {\it dynamics} of
strong glass formers~\cite{horbach96} and also the DOS (see below)
shows strong finite size effects which have to be avoided.

The equations of motion have been integrated in the microcanonical
ensemble with the velocity form of the Verlet algorithm using a time
step of 1.6 fs. To generate the glass we proceeded as follows:  First
we equilibrated the system for 4 million time steps at a relatively
high temperature (2900~K). At this temperature the melt is already
quite viscous (120 Poise)~\cite{horbach_diss,kob98} and the typical
relaxation times are of the order of 3~ns~\cite{horbach98a}.
Subsequently the system was coupled to an external heat bath whose
temperature was decreased linearly in time for 1.0  million time steps
to zero Kelvin. This corresponds to a cooling rate of about 1.8$\cdot
10^{12}$~K/s. With this cooling rate the system falls out of equilibrium
at around 2850~K and thus this temperature corresponds to about the
value of the fictive temperature of the glass. During the cooling
procedure we stored periodically copies of the sample. These copies
were then used as starting configurations for a run in the canonical
ensemble with 500,000 time steps (0.8 ns) in order to anneal the
configurations.  Afterwards (microcanonical) runs were started in which
the time Fourier transform of the velocity autocorrelation function was
measured which was subsequently used to calculate the density of
states. In order to improve the statistics of the results we averaged
over four independent runs.

\section{Results}
\label{sec3}

In order to calculate the specific heat within the harmonic
approximation we need the density of states (DOS), $g(\nu)$. To obtain
$g(\nu)$ one can, e.g., proceed as follows: For any given configuration
of particles one uses a steepest descent method to determine the
location of the nearest (metastable) minimum of the potential energy.
The DOS can then be calculated from the eigenvalues of the
(mass-weighted) Hessian matrix at this local minimum. This approach
was, e.g., used in Ref.~\cite{vollmayr96b} to determine how the DOS
depends on the cooling rate with which the sample was cooled from the
high temperature phase to temperatures below $T_g$. Below we will come
back to these results.

A different method to obtain the DOS is to determine the time Fourier
transform of the velocity autocorrelation function and to make use of
the relation~\cite{dove93}

\begin{equation}
g(\nu) = \frac{1}{Nk_BT}\sum_j m_j 
\int_{-\infty}^{\infty} dt \exp(i2\pi\nu t) \langle \vec{{\bf v}}_j(t)
\vec{{\bf v}}_j(0) \rangle \quad .
\label{eq2}
\end{equation}

It should be noted that this approach gives an {\it effective} DOS
which, however, coincides at low temperatures, i.e. when the harmonic
approximation is valid, with the real one.

\begin{figure}[f]
\psfig{file=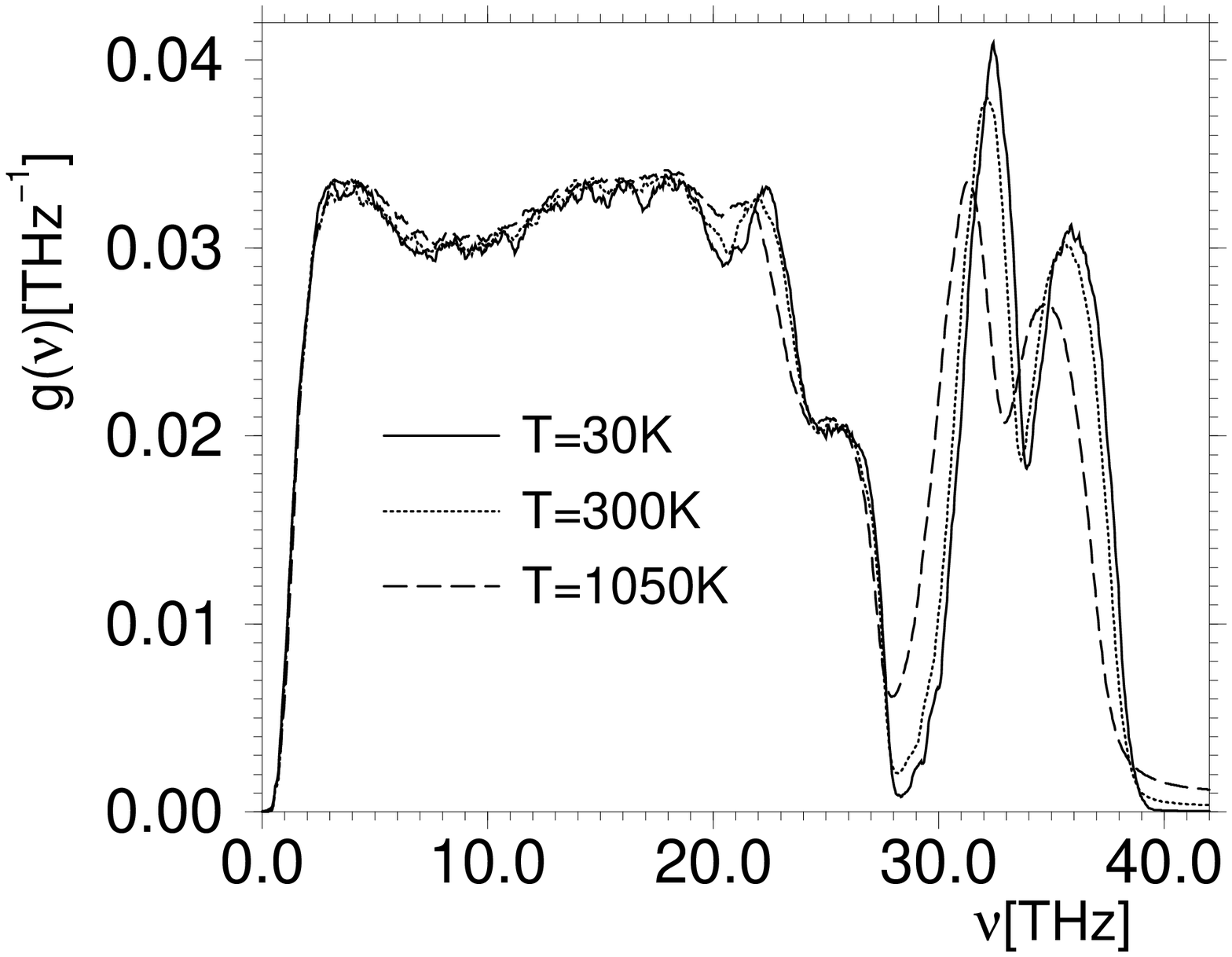,width=13cm,height=9.5cm}
\caption{Effective density of states for $T=$30K (solid line),
$T=300$K (dotted line) and $T=1050$K (dashed line).}
\label{fig1}
\end{figure}

In Fig.~\ref{fig1} we show this effective DOS for three different
temperatures, $T=30$~K, $T=300$~K, and $T=1050$~K. Note that all of
these temperatures are well below the glass transition
temperature~($T_g$=1450K) and thus no relaxational processes take
place. From the figure we recognize that the main feature of the DOS is
a double peak at high frequencies and a relatively flat region at
intermediate frequencies, as already discussed in
Refs.~\cite{taraskin97b,vollmayr96b,valle94}.  It has been shown that
the two peaks at high frequencies are due to intra-tetrahedral motions
of the ions whereas the frequencies of the inter-tetrahedral motions
are in the broad region of $g(\nu)$ at lower
frequencies~\cite{galeener79,pasquarello98}. From the temperature
dependence of the DOS we see that the main change in $g(\nu)$ occurs at
high frequencies in that, with decreasing temperature, the height of
the two peaks increases and that their location shifts to higher
frequencies, i.e. that the corresponding vibrations become faster. (We
also note that the curve for $T=1050$K shows a tail which extends
beyond 40THz. We have checked that this feature is not an effect of the
sampling procedure and thus conclude that it originates from the
anharmonicity of the system.) At intermediate and low frequencies
the temperature dependence of the DOS is much weaker. All these
observations are in qualitative agreement with the results of Vollmayr
{\it et al.} where a similar dependence was found when the glass
transition temperature $T_g$ was varied~\cite{vollmayr96b}. It should
be noted, however, that in that work the dependence of $g(\nu)$ on
$T_g$ was related to the fact that the structure of the system changes
when the cooling rate is varied. In contrast to this we investigate
here only one cooling rate and the temperature dependence of the
effective DOS stems only from anharmonic effects and not from a
structural change.

We also note that at very small frequencies the DOS shows a gap. As
already discussed in Ref.~\cite{vollmayr96b} this is a finite size
effect in that excitations which have a spatial extension larger than
the size of the simulation box are not present in this system. Some
of these excitations are, e.g., acoustic phonons with a large wave
length. In the Debye theory the density of states of these phonons
is given by 

\begin{equation}
g_D(\nu)=3\nu^2/\nu_D^3
\label{eq3}
\end{equation}

where the Debye frequency $\nu_D$ is related to $c_t$ and $c_l$, 
the transverse and longitudinal speed of sound, by

\begin{equation}
\nu_D=\left( \frac{9N}{4\pi V}\right)^{1/3} \left(
\frac{2}{c_t^3}+\frac{1}{c_l^3} \right) ^{-1/3}.
\label{eq4}
\end{equation}

Here $N$ is the number of particles in the volume $V$. By measuring the
dispersion relation of the transversal and longitudinal acoustic waves
at small wave-vectors, we have determined the values of $c_t$ and
$c_l$~\cite{horbach98b,horbach_diss}. We have found that these values,
$c_t=3772$m/s and $c_l=5936$m/s, agree very well with the experimental
values $c_t=3767$m/s~\cite{malinovsky86} and
$c_l=5970$m/s~\cite{benassi96}, thus demonstrating that with respect to
these quantities our model is quite realistic\footnote{Note that these
figures are valid for 300~K. Also it has to be taken into account,
that the density of our system was fixed at 2.37g/cm$^3$, which is
slightly higher than the experimental value 2.2g/cm$^3$. Hence our
sound velocities have to be multiplied by $\sqrt{2.37/2.2}$ before they
are compared with the experimental values. If this is done one obtains
$c_t=3915$m/s and $c_l=6161$m/s.}. For $\nu_D$ we obtain 10.65THz which
compares well with the experimental value which is
10.40THz~\cite{vacher81}. Therefore we have closed the mentioned gap in
our DOS by adding to our measured $g(\nu)$ the Debye-law given by
Eq.~(\ref{eq3}) in the frequency range $0\leq \nu \leq 0.73$THz. We
will show below that this modification leads to a significant
improvement of the specific heat at low temperatures.

\begin{figure}[h]
\psfig{file=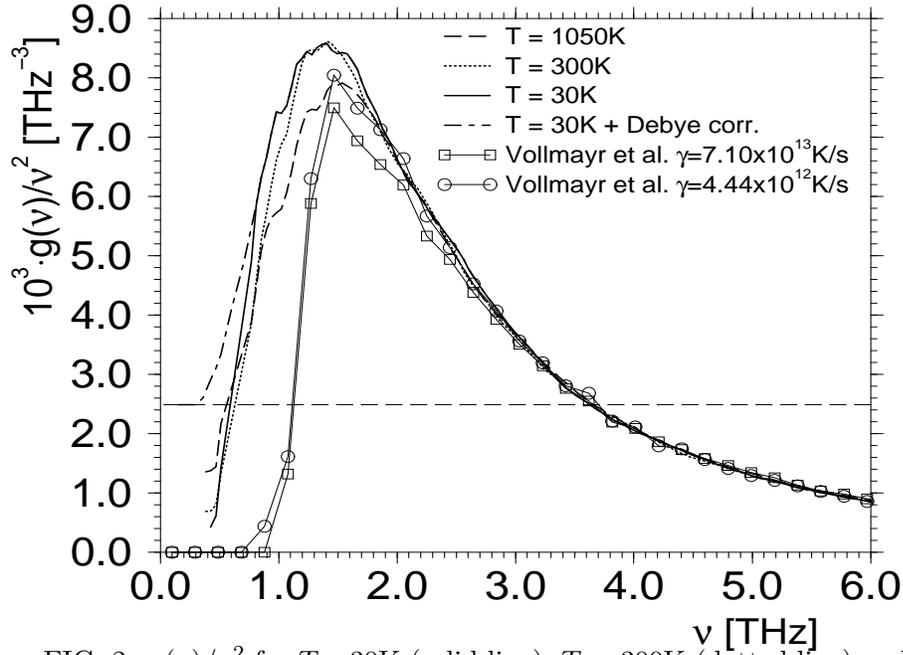,width=12cm,height=9.0cm}
\caption{$g(\nu)/\nu^2$ for $T=$30K (solid line),
$T=300$K (dotted line) and $T=1050$K (dashed line). The dashed-dotted
curve is the $T=30$K curve including the Debye corrections. The two
curves with symbols are from the simulation of Vollmayr {\it et al.}
~\protect\cite{vollmayr96b} for a $N=1002$ ion system. The horizontal
dashed line is the Debye value.}
\label{fig2}
\end{figure}

At low temperatures the temperature dependence of the specific heat
will be dominated by the low frequency part of the spectrum. This
frequency range is also of particular interest because it includes the
so-called boson peak, a dynamical feature whose nature is currently a
matter of intense
debate~\cite{buchenau86,taraskin97b,benassi96,bp_papers}. As mentioned
above, at low frequencies the DOS is expected to show the mentioned
Debye behavior, i.e. $g(\nu)\propto \nu^2$.  In Fig.~\ref{fig2} we
therefore plot $g(\nu)/\nu^2$ for the three different temperatures
(bold lines). We see that the DOS shows a large peak at around 1.4THz,
i.e. an excess DOS over the Debye value (horizontal dashed line), the
mentioned boson peak~\cite{buchenau86,taraskin97b,benassi96,bp_papers}.
Comparing the curves for the three different temperatures, we see that
in this frequency range the harmonic approximation holds up to
temperatures of 300K, since only the curve for $T=1050$K (dotted line)
differs significantly from the ones at the two lower temperatures. Note
that the curves shown are the one without the mentioned Debye
correction. In order to see the effect of this correction we include
the DOS for $T=30$K when the correction is taken into account
(dash-dotted curve). We see that, as expected by construction, this
corrected curve goes for small frequencies to the Debye value and joins
the measured DOS at 0.73THz.

Also included in the figure are two curves from the work of Vollmayr
{\it et al.}~\cite{vollmayr96b}. These were obtained by cooling a
sample of $N=1002$ ions with two different cooling rates $\gamma$,
namely $\gamma=7.1\cdot 10^{13}$K/s and $\gamma=4.4\cdot 10^{12}$K/s, to
zero temperature and calculating the DOS from the dynamical matrix.
From these two curves we recognize that the height of the boson-peak
depends on the cooling rate, as already discussed in
Ref.~\cite{vollmayr96b}, and that the low frequency side of the peak of the
two curves is at a higher frequency than the one of the curves for 8016
ions. One might be tempted to add the mentioned Debye correction also
to the curves for the smaller system, but a look at the magnitude of
these corrections shows that they are not sufficient to bring the
curves for $N=1002$ in coincidence with the ones of $N=8016$. Thus we
conclude that the excitations giving rise to the boson peak do have a
significant spatial extension and thus it is necessary to use large
system sizes if finite size effects have to be avoided.  

From the DOS it is easy to calculate the temperature dependence of
$C_V$, the specific heat at constant volume. It is given by

\begin{equation}
C_V=\frac{h^2}{k_BT^2}\int_0^\infty \frac{\nu^2
\exp(h\nu/k_BT)}{\left( \exp(h\nu/k_BT)-1\right)^2} g(\nu) d\nu.
\label{eq5}
\end{equation}

\begin{figure}[h]
\psfig{file=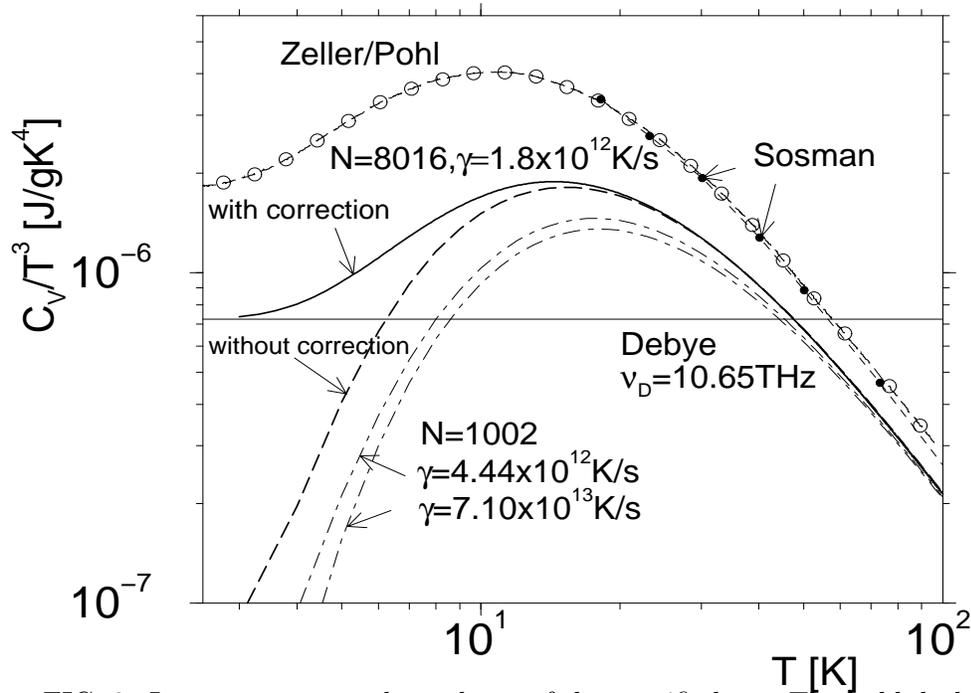,width=13cm,height=9.0cm}
\caption{Low temperature dependence of the specific heat. The bold
dashed curve is $C_V/T^3$ as calculated from the DOS at 30K and the
bold curve is obtained if the Debye corrections are taken into account.
The curves with symbols are the experimental results by Sosman (filled
circles) and by Zeller and Pohl (open squares)~\protect\cite{zeller_sosman}.
The dashed-dotted lines are the ones obtained from the DOS of
Vollmayr {\it et al.} for two different cooling 
rates~\protect\cite{vollmayr96b}. The horizontal line is the Debye
law.}
\label{fig3}
\end{figure}

At low temperatures the specific heat is expected to show a Debye
law, i.e. $C_V \propto T^3$, and therefore it is useful to plot
$C_V/T^3$ vs. $T$, which is done in Fig.~\ref{fig3}. The bold dashed
curve is the specific heat as obtained from Eq.~(\ref{eq5}) using the
measured DOS at $T=30$K and the solid bold curve is $C_V$ from this
DOS and the Debye corrections. The horizontal dashed line is the
Debye law 

\begin{equation}
C_V(T)=\frac{12\pi^4Nk_B}{5}\left(\frac{T}{\Theta_D}\right)^3 \quad,
\label{eq6}
\end{equation}

with the Debye temperature $\Theta_D=h\nu_D/k_B$, which for our model
is $\Theta_D=511$K. We see that the difference between the uncorrected
and corrected curve becomes notiecable for temperatures below 20K in
that the former decreases towards zero for $T\to 0$ whereas the latter
approaches the Debye value given by Eq.~(\ref{eq6}). 

Also included in the figure is experimental data by Sosman and by
Zeller and Pohl~\cite{zeller_sosman}. We see that although the
corrected curve looks qualitatively similar to the experimental data,
the location of the peak is overestimated by about 5~K.  Furthermore
also the height of the peak, as well as the other parts of the
theoretical curve, underestimates the measured values by about a factor
of two, a result which is in agreement with findings of Taraskin and
Elliott~\cite{taraskin97b}. This discrepancy is not due to the presence
of two-level systems, since these are relevant only at lower
temperatures~\cite{two_level}.  Also shown are two curves that stem
from the DOS calculated by Vollmayr {\it et al.} (see Fig.~\ref{fig2}).
A comparison of these two curves with the uncorrected one from the
present simulation shows that there are at least two possible reasons
for the observed discrepancy between the theoretical and experimental
curves for the specific heat. First of all we see from the curves for
$N=1002$ ions that there is a small but noticable cooling rate
dependence of $C_V(T)$ in that the specific heat increases with
decreasing cooling rate. Thus it can be expected that if our silica
model would be cooled with a typical experimental cooling rate
[$O$(1K/s)], the theoretical curve would be significantly higher than
the one shown in the figure. Secondly we see that the theoretical
curves also show a substantial system size dependence in that the one
for the larger system is, at low temperatures, significantly above the
one for the smaller system. (The fact that the curve for the larger
system has a cooling rate which is by a factor of two smaller than the
one for the smaller cooling rate of the smaller system is not
sufficient to explain this discrepancy between the curves for the two
system sizes.) Hence we conclude that the anomalous behavior of the
specific heat is also considerably affected by finite size effects.

\begin{figure}[h]
\psfig{file=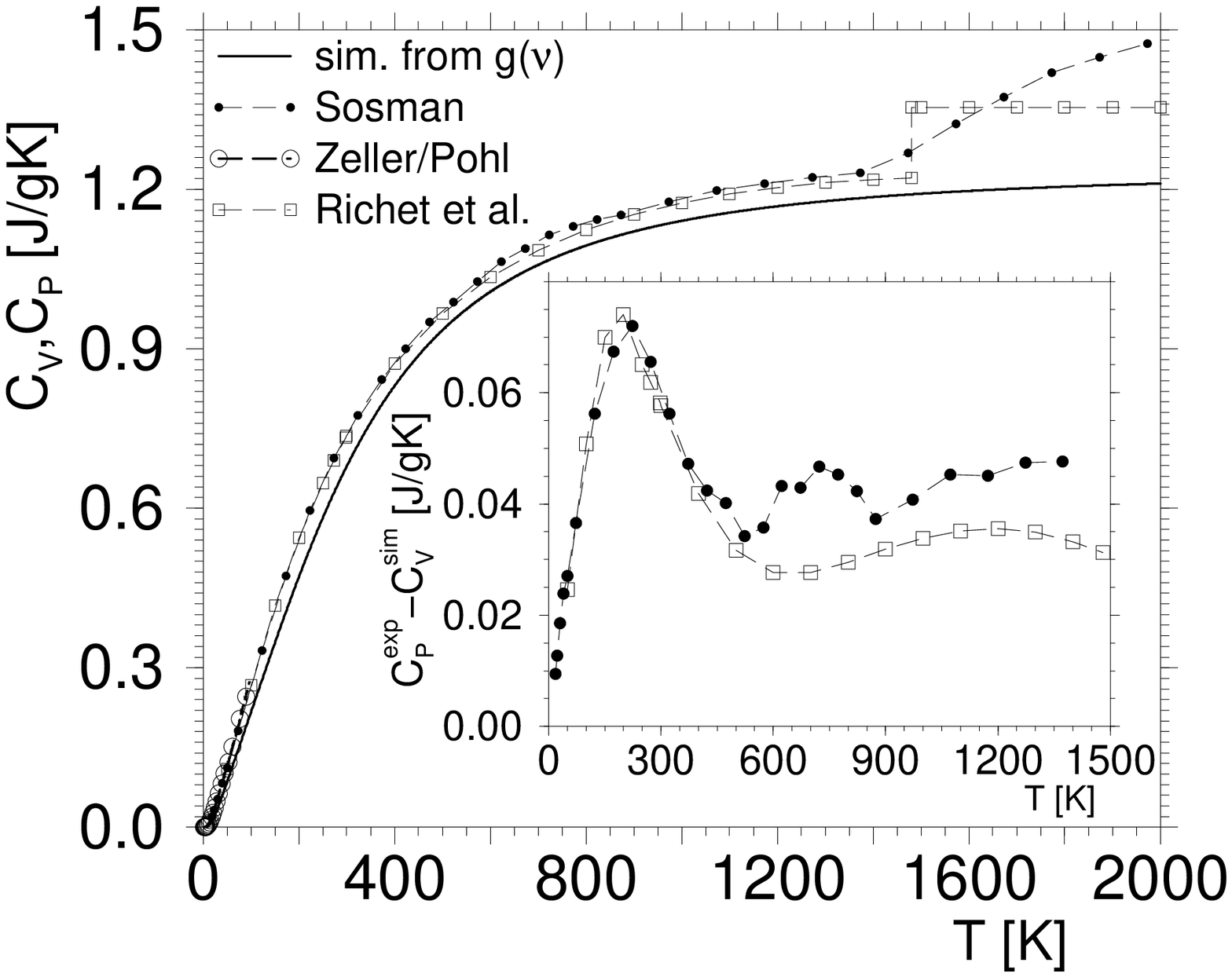,width=13cm,height=9.0cm}
\caption{Temperature dependence of the specific heat at constant volume
as predicted by the harmonic approximation (solid line). The symbols
are experimental data for the specific heat at constant pressure by
Sosman, by Zeller and Pohl, and by Richet {\it et
al.}~\protect\cite{zeller_sosman,richet82}. Inset:  Difference between
the data of Richet {\it et al.} and Sosman and our data.}
\label{fig4}

\end{figure}

From the figure we also see that with increasing temperature the
relative difference between the theoretical and the experimental curves
decreases. Therefore it is interesting to compare these curves also at
higher temperatures, which is done in Fig.~\ref{fig4}. Here we show
$C_V(T)$, as calculated from our DOS at 30K (solid line), and $C_P(T)$,
the specific heat at constant pressure, as determined by various
experiments~\cite{zeller_sosman,richet82} (lines with symbols).  We
recognize that for temperatures below the glass transition temperature
$T_g=1450$K the agreement between our calculation and the experimental
data is very good in that the difference is smaller than 0.05 J/gK (see
inset). At $T_g$ the theoretical curve starts to deviate from the
experimental data since at this temperature the real system starts to
flow and thus the translational degrees of freedom associated with this
motion contribute to the specific heat. Since the harmonic model does
not include this type of motion the theoretical curve does not show
any special temperature dependence at $T_g$ and with increasing
temperature approaches the classical value of Dulong and Petit,
1.236J/gK.

Since the experimental data shown in the figure stem from
measurements at constant pressure, one has to check whether the
discrepancy between the experimental and theoretical curves is
due to the difference between $C_V$ and $C_P$. 
This difference can be expressed by the thermodynamic relation
\begin{equation}
C_P-C_V=TV\frac{\alpha^2}{\kappa_T} \quad,
\label{eq7}
\end{equation}
where $\alpha$ is the thermal expansion coefficient and $\kappa_T$ is
the isothermal compressibility. Using the experimental values
$\alpha=5.5 \cdot 10^{-7}$K$^{-1}$, valid in the temperature range 293K
$\leq T \leq $ 593K~\cite{alpha_ref}, and $\kappa_T=2.79 \cdot
10^{-5}$GPa$^{-1}$ at $T=1173$K~\cite{bruckner70}, one obtains an
estimate for $C_P-C_V \approx 5 \cdot 10^{-3}$J/gK. This is about one
decade less than the observed difference between our theoretical values
and the experimental data and hence we conclude that the observed
(small) discrepancy is due to some inadequacy of the model. (We also
mention that from the temperature dependence of the static structure
factor at small wave-vectors we have shown that the compressibility of
our system is in reasonable good agreement with experimental values,
therefore giving support to the above
estimate~\cite{horbach_diss,horbach99}. Finally we mention that the
observed discrepancy also does not disappear if instead of the DOS for
30~K we use the one at 300~K or at 1050~K. Hence the difference between
the theory and the experiment is indeed due to the used model for
silica.

\section{Conclusions}
\label{sec4}

We have presented results from a molecular dynamics computer simulation
of amorphous silica in which we used the velocity autocorrelation
function to calculate an effective density of states $g(\nu)$. We find
that for temperatures less than around 300~K $g(\nu)$ is essentially
independent of $T$, i.e. that the harmonic approximation is valid,
whereas significant deviations are observed for $T=1050$~K. Since this
last temperature is still significantly below the experimental galss
transition temperature ($T_g$=1450~K), and even further below the glass
transition temperature of this simulation ($T_{g,{\rm sim}}\approx
2850$K) we do not expect that the system relaxes on the time scale of
the simulation. Hence the temperature dependence of $g(\nu)$ is due to
the anharmonic nature of the local potential wells in which the ions
are sitting.

We demonstrate that for frequencies in the vicinity of the
boson-peak, i.e. around 1-2THz, the DOS shows a quite strong dependence
on the cooling rate, and more important on the system size. Hence it
can be concluded that the excitations giving rise to the boson-peak
have a relatively large spatial extention so that even our simulation
box, which measures about 48\AA, is not able to include all of them.  Because
of this limited system size the low-temperature specific heat we obtain
from our simulation is about a factor of two smaller than the one found
in experiments. For higher temperatures we find, however, that our
theoretical curves agree very well with the experimental data, thus
showing that this type of calculation is surprisingly reliable for
temperatures below $T_g$.

Acknowledgements: We thank Dr. U. Fotheringham for suggesting the
calculations presented in this work. This work was supported by BMBF
Project 03~N~8008~C and by SFB 262/D1 of the Deutsche
Forschungsgemeinschaft.  We also thank the HLRZ J\"ulich for a generous
grant of computer time on the T3E.

\clearpage
\newpage

\vspace*{-5mm}
\par

\begin{references}
\bibitem{buchenau86}
U. Buchenau, M. Prager, N. N\"ucker, A. J. Dianoux, N. Ahmad, and W.
A. Phillips,
Phys. Rev. B {\bf 34}, 5665 (1996).

\bibitem{taraskin97b}
S. N. Taraskin and S. R. Elliott,
Phys. Rev. B {\bf 56}, 8605 (1997).

\bibitem{beest90}
B. W. H. van Beest, G. J. Kramer, and R. A. van Santen,
Phys. Rev.  Lett. {\bf 64}, 1955 (1990).

\bibitem{vollmayr96b}
K. Vollmayr, W. Kob, and K. Binder,
Phys. Rev. B {\bf 54}, 15808 (1996).

\bibitem{vollmayr96a}
K. Vollmayr and W. Kob,
Ber. Bunsenges. Phys. Chemie {\bf 100}, 1399 (1996).

\bibitem{horbach98a}
J. Horbach, W. Kob, and K. Binder,
Phil. Mag. B {\bf 77}, 297 (1998).

\bibitem{horbach98b}
J. Horbach, W. Kob, and K. Binder,
J. Non-Cryst. Solids {\bf 235-238}, xxxx (1998).

\bibitem{horbach96}
J. Horbach, W. Kob, K. Binder, and C. A. Angell,
Phys. Rev. E. {\bf 54}, R5897 (1996).

\bibitem{horbach_diss}
J. Horbach, PhD Thesis, University of Mainz, (1998).

\bibitem{kob98}
W. Kob and K. Binder,
to appear in {\it Analysis of Composition and Structure of Glass and
Glass Ceramics,} Eds.: H. Bach and D. Krause (Springer, Berlin, 1998).

\bibitem{dove93}
M. T. Dove, {\it Introduction to Lattice Dynamics} (Cambridge
University Press, Cambridge, 1993).

\bibitem{valle94}
R. G. Della Valle and E. Venuti, Chem. Phys. {\bf 179}, 411 (1994).

\bibitem{galeener79}
F. L. Galeener,
Phys. Rev. B {\bf 19}, 4292 (1979).

\bibitem{pasquarello98}
A. Pasquarello and R. Car,
Phys. Rev. Lett. {\bf 80}, 5145 (1998).

\bibitem{malinovsky86}
V. Malinovsky and A. P. Sokolov,
Sol. State Comm. {\bf 57}, 757 (1986).

\bibitem{benassi96}
P. Benassi, M. Krisch, C. Masciovecchio, V. Mazzacurati, G. Monaco,
G. Ruocco, F. Sette, and R. Verbeni,
Phys. Rev. Lett. {\bf 77}, 3835 (1996).

\bibitem{vacher81}
R. Vacher, J. Pelous, F. Plicque, and A. Zarembowitch, J.
Non-Cryst. Solids {\bf 45}, 397 (1981).

\bibitem{bp_papers}
M. Foret, E. Courtens, R. Vacher, and J.-B. Suck,
Phys. Rev. Lett.  {\bf 77}, 3831 (1996);

W. Schirmacher, G. Diezemann, and C. Ganter,
Phys. Rev. Lett. {\bf 81}, 136 (1998);

A. Wischnewski, U. Buchenau, A. J. Dianoux, W. A. Kamitakahara, and
J. L. Zarestky,
Phys. Rev. B {\bf 57}, 2663 (1998).

\bibitem{zeller_sosman}
R. B. Sosman, {\it The Properties of Silica} (Chemical Catalog Company,
New York, 1927);
%
R. C. Zeller and R. O. Pohl, Phys. Rev. B {\bf 4}, 2029 (1971).

\bibitem{two_level}
W. A. Phillips, J. Low Temp. Phys. {\bf 7}, 351 (1972);
%
P. W. Anderson, B. I. Halperin, and C. M. Varma, Phil. Mag. {\bf 25},
1 (1972).

\bibitem{richet82}
P. Richet, Y. Bottinga, D. Denielou, J. P. Petitet, and C. Tegui,
Geochim. Cosmochim. Acta {\bf 46}, 2639 (1982).

\bibitem{alpha_ref}
{\it Handbook of Chemistry and Physics}, Eds. R. C. Weast, M. J.
Astle, and W. H. Beyer (The Chemical Rubber Co., Boca Raton, 1984).

\bibitem{bruckner70}
R. Br\"uckner, J. Non-Cryst. Solids  {\bf 5}, 123 (1970).

\bibitem{horbach99}
J. Horbach, W. Kob, and K. Binder, to be published.

\end{references}
\end{document}